# A frame semantic overview of NLP-based information extraction for cancer-related EHR notes

Surabhi Datta[1], Elmer V Bernstam[1,2], Kirk Roberts[1]
[1]School of Biomedical Informatics, The University of Texas Health Science Center at Houston, TX USA
[2]Department of Internal Medicine, McGovern Medical School, The University of Texas Health Science Center at Houston, TX USA

Corresponding Author:
Kirk Roberts, PhD
7000 Fannin St #600
Houston TX 77030
kirk.roberts@uth.tmc.edu

**ABSTRACT**

*Objective*: There is a lot of information about cancer in Electronic Health Record (EHR) notes that can be useful for biomedical research provided natural language processing (NLP) methods are available to extract and structure this information. In this paper, we present a scoping review of existing clinical NLP literature for cancer.

*Methods*: We identified studies describing an NLP method to extract specific cancer-related information from EHR sources from PubMed, Google Scholar, ACL Anthology, and existing reviews. Two exclusion criteria were used in this study. We excluded articles where the extraction techniques used were too broad to be represented as frames (e.g., document classification) and also where very low-level extraction methods were used (e.g. simply identifying clinical concepts). 79 articles were included in the final review. We organized this information according to frame semantic principles to help identify common areas of overlap and potential gaps.

*Results*: Frames were created from the reviewed articles pertaining to cancer information such as cancer diagnosis, tumor description, cancer procedure, breast cancer diagnosis, prostate cancer diagnosis and pain in prostate cancer patients. These frames included both a definition as well as specific frame elements (i.e. extractable attributes). We found that cancer diagnosis was the most common frame among the reviewed papers (36 out of 79), with recent work focusing on extracting information related to treatment and breast cancer diagnosis.

*Conclusion*: The list of common frames described in this paper identifies important cancer-related information extracted by existing NLP techniques and serves as a useful resource for future researchers requiring cancer information extracted from EHR notes. We also argue, due to the heavy duplication of cancer NLP systems, that a general purpose resource of annotated cancer frames and corresponding NLP tools would be valuable.

**INTRODUCTION**

Unstructured clinical data about cancer patients is increasingly available in clinical notes in Electronic Health Records (EHRs) and other systems. There is increasing interest in utilizing this data for biomedical research [1][2] [3][4][5][6][7][8]. Furthermore, with the emergence of precision medicine, the need to extract more and more detailed information from a patient's EHR notes becomes ever greater. Many publications argue for the need to capture important clinical information on cancer, ranging from information about tissue specimens [9] to disease-related and outcome information [10], thus facilitating translational research by associating molecular information with disease phenotype [11]. Many of these information types are found nearly exclusively in unstructured or semi-structured format in EHRs. For example, information associated with biomarkers and cancer prognosis are frequently stored in free-text surgical pathology reports [12]. Some work focuses on extracting information to improve cancer screening efficiency [1] [7][13][14][15]. All such work has focused on deriving cancer-related information automatically using various natural language processing (NLP) techniques. Although a majority of the methods applying deep learning in cancer research are for processing images (e.g. mammograms) [16][17][18][19] and gene expression profiles [20][21][22], more recently, deep-learning based NLP systems are gaining prominence for cancer information extraction from EHRs [23][24][25] [26][27]. For example, Gao et al. [26] implemented a hierarchical attention network for extracting some of the crucial clinical oncology data elements such as primary cancer site and histological grade which are gathered by cancer registries.

However, many of these researchers put sizable effort into designing and implementing NLP systems that extract similar information types. In this scoping review, we investigate which cancer information types have been extracted with NLP techniques. We organize the extracted information into frames, based on the linguistic theory of 'frame semantics'. Frames provide a convenient, flexible, and linguistically-motivated representation for information as complex and diverse as that related to cancer. Our aim is to provide a list of cancer-related frames in existing work that would be valuable to the scientific community.

Frame semantics is a linguistic theory that postulates the meaning of most words is understood in relation to a conceptual frame in which entities take part. E.g., the meaning of sell in the "*Jerry* *sold a car to Chuck*" evokes a frame related to COMMERCE, which includes four elements: BUYER, SELLER, MONEY, and GOODS, though not all elements are required (as with MONEY here). Frames can represent all events types, from the simple PLACING frame ("*Maria put money in her account*") with elements AGENT, THEME, and GOAL, to the more complex HIRING ("*He hired Downing as his coach in Hawai'i*") with elements EMPLOYER, EMPLOYEE, POSITION, and PLACE. Frames can also encode relations (e.g., KINSHIP: "*Joe's brother John*") and states (e.g., BEING_IN_OPERATION: "*the centrifuge is operational*"). Currently, the Berkeley FrameNet database [28] contains more than 1,200 frames with an average of about 10 elements per frame. Manual annotations of more than 200,000 frame instances provide a unique level of detail about how these elements can appear in sentences. The frames are connected by more than 1,800 frame relations, forming a lattice structure. As exemplified by the above examples, however, FrameNet has minimal coverage of biomedical information. This review may be viewed, then, as part of a proposed set of frames specifically targeting cancer information in EHR notes. On the other hand, frame semantics simply provides a useful mechanism for organizing the wide variety of information found in the papers covered by this review. We make no claim these frames are the 'best' way to organize cancer information for non-NLP purposes. These frames are, however, a natural endpoint for a scoping review, as they help to succinctly define the scope of cancer information that has been extracted from EHR free text.

**MATERIALS AND METHODS**

The goal of this study is to review papers on NLP for EHR notes related to cancer, determine what (parts of) frame(s) the researchers aim to extract, then express the complete set of frames in a consistent way across all the identified papers. The focus is thus the type of cancer data that is extracted using NLP, rather than on the NLP method used or its performance. PubMed and Google Scholar were searched using the keywords 'natural language processing' or 'NLP' anywhere in the article and one of the three keywords from 'cancer', 'tumor', and 'oncology' in the title. We also searched the ACL Anthology using only the cancer related keywords ('cancer', 'tumor', or 'oncology') in the title. Results were limited to papers published after January 1, 2000 in order to capture the current generation of statistical and machine learning-based NLP (though we did not exclude articles after this date based on method). Additionally, citations from two cancer NLP reviews [29][30] were used. Finally, relevant citations in the reviewed papers were iteratively added to the review process.

We obtained a total of 899 articles using the above search process (108 via PubMed, 738 via Google Scholar, 16 via ACL Anthology, and 37 via the existing reviews). After de-duplication, 703 articles were chosen for title/abstract screening.

**Screening Process**

We selected relevant articles based on whether the title or abstract contained the description of an NLP method to extract cancer-related information from EHR notes, or suggests cancer-related information is extracted from free text EHR sources. Two raters (SD, KR) independently assessed whether the title or abstract suggested that the data source was not related to EHR notes (e.g., literature [31], social media [32][33]). Papers describing NLP on data sources other than EHR notes were excluded. If the data source was not clear, or the raters could not agree during reconciliation, the paper was kept for full-text review.

For the 173 papers that passed the initial title/abstract screening, two further exclusion criteria were applied to the full text. First, document-level text classification methods were excluded. Not only are these not considered information extraction NLP methods and are not amenable to frame representation, they are often application specific and not re-purposable. For example, an article by Garla et al. [34] describing a binary classifier to determine whether a clinical report is about a potential malignant liver lesion requiring follow up would not be considered for this study. This is a very common use case of NLP for cancer research, but these methods do not attempt to identify specific cancer-related data elements, and are thus not amenable to frames. On the other hand, an article that extracts references to malignant liver lesions and factors (e.g., phrases) that may impact the need for follow up would be relevant for this study since those factors could be organized by frame. Second, articles were excluded where very unspecific extraction methods were used like concept recognition and named entity recognition techniques (such as the use of MetaMap [35] or cTAKES [36]) that identify phrases but do not attempt to connect these to their wider context [37][38][39]. For example, an article by Xie et al. [39] that does not differentiate between an asserted and a negated concept, such as extracting the phrase

"breast cancer" regardless of the context being "has breast cancer" vs. "has no breast cancer", is not sufficiently semantically expressive and would thus be excluded. Similarly, an article describing the use of a general-purpose concept extraction tool on mammography reports would not be considered relevant for frame representation. These types of NLP articles can be viewed as describing the basic building blocks upon which a frame-based information extraction system can be built, but they do not capture sufficient contextual information to be properly described as frames. In cases where the exclusion decision could not be made from the title and abstract alone, the full text was screened. In total, title/abstract screening resulted in removing 530 articles and the full-text review of the remaining 173 articles eliminated a further 94 articles. Finally 79 articles were included in this scoping review. Figure 1 shows the overall process including numbers removed at each stage.

**Frame Construction**

For each of the 79 articles relevant to frame semantics for cancer in EHR notes, one or more frames were constructed to represent the information types extracted by the described NLP system. Note that NLP systems are not typically described in a frame semantic manner nor use the terminology of frames. Instead, this had to be inferred based on the descriptions. For papers that focus on one or two data types, this is a straightforward process that results in a single, small frame description or a single element within a larger frame. For papers that focus on multiple data types (at least 7 papers describe over 5 different data types), the data types were organized into one or more frames. Multiple frames were used when the information was conceptually different (e.g., the description of a tumor versus the description of a cancer procedure). However, this was an iterative, collaborative process: batches of 20 papers were considered at a time, and two reviewers proposed initial frames for each of the papers. Disagreements were reconciled, after which the next batch was considered, which may have resulted in new frames as well as modifications to frames created in previous batches. A common set of frames was created that described important cancer-related information that NLP techniques have been used to extract.

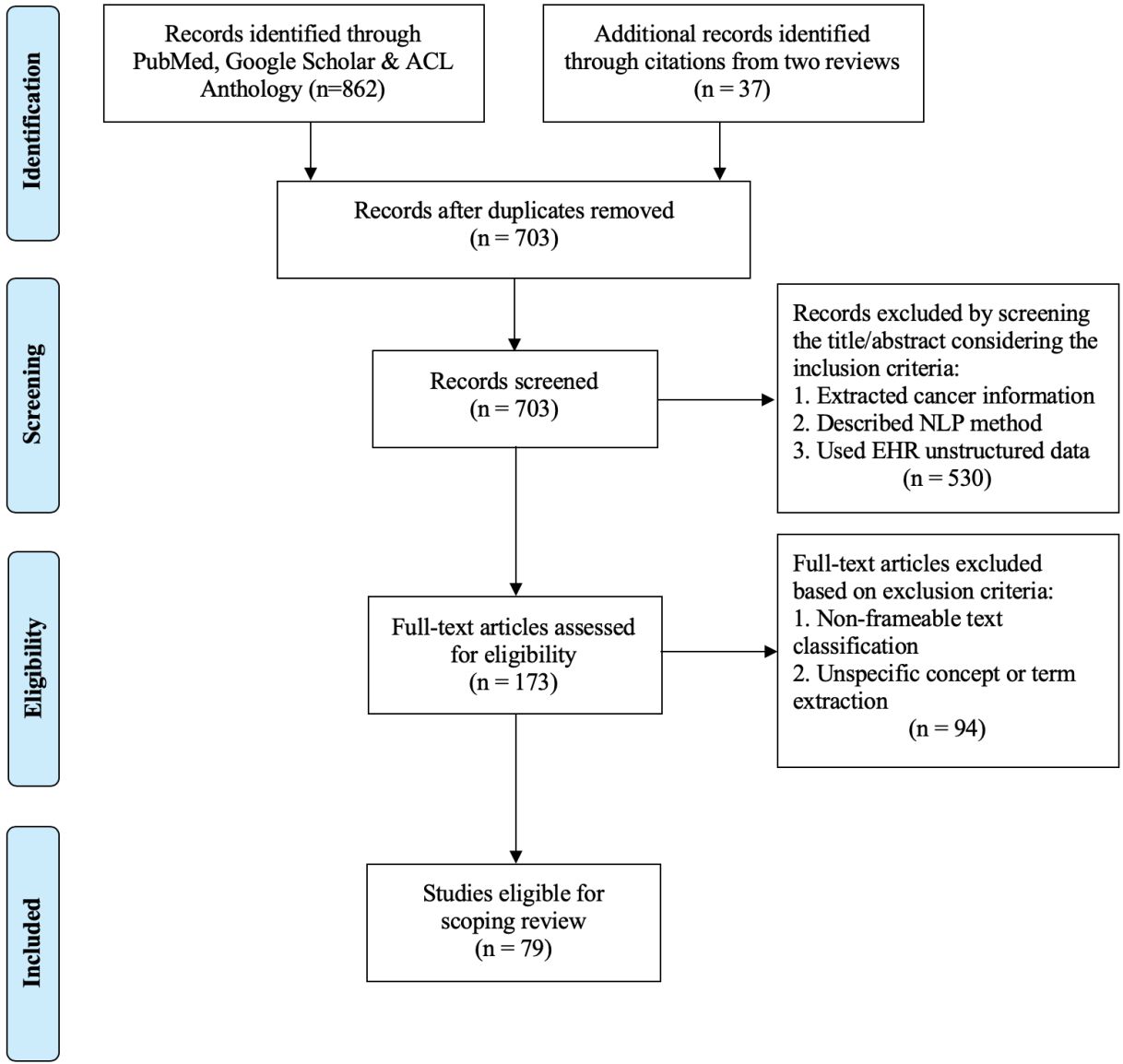

**Figure 1: PRISMA diagram for study selection**

## RESULTS

**Frame Descriptions**

Each constructed frame has a name along with a definition (see Figure 2). Moreover, each frame is described using some attributes or elements, known as 'frame elements'. The frame elements correspond to the specific items extracted by one or more NLP system included in this study. Each of these elements is defined briefly along with identifying the corresponding articles that extract that frame element (Table 1). For example, 'CANCER DIAGNOSIS' is a frame, which is defined as 'an event of a patient being diagnosed with a cancer', and described using frame elements such as ANATOMICAL SITE, HISTOLOGY, and DIAGNOSIS STATUS.

**Frame Relations**

The constructed frames additionally have relations between them based on those found in Berkeley FrameNet [28]. First is the 'parent-child' or 'inheritance' relation ('inherits' in Figure 2). More specifically, every child frame is a specific version of the parent frame. For example, 'CANCER DIAGNOSIS' frame is inherited by multiple frames such as 'BREAST CANCER DIAGNOSIS' and 'BLADDER CANCER DIAGNOSIS'. The elements in 'CANCER DIAGNOSIS' frame are generic to diagnosis of any cancer type, whereas 'BREAST CANCER DIAGNOSIS' frame contains extra breast cancer-specific elements such as RECEPTOR STATUS and RECEPTOR PERCENT. Similarly, an element such as MUSCULARIS PROPRIA is more prevalent in bladder cancer diagnosis cases.

The second relation is 'an element of' where a frame is an element described in another frame. This is analogous to the 'Subframe' relation proposed in the Berkeley FrameNet project. As mentioned in Table 1, TNM CLASSIFICATION is an element present in both 'CANCER DIAGNOSIS' and 'TUMOR DESCRIPTION' frames, while at the same time, 'TNM CLASSIFICATION' is described as a separate frame altogether. A separate frame for a frame element was created if the element has multiple attributes and at least one included paper extracts such attributes. For example, AAlAbdulsalam et al. [2] extracts elements which are very specific to TNM stage such as the actual stage designation (e.g. T1, N1b), staging method (e.g. clinical or pathological staging), and negation. This information can be better understood if a separate frame 'TNM CLASSIFICATION' is created to contain TNM staging related details.

The third frame relation is 'associated with', which indicates any kind of association or dependency between two frames. For example, the 'CANCER FINDING' frame is associated with the 'THERAPEUTIC PROCEDURE' frame as the findings in clinical reports often aid in selecting the best therapy for treating cancer patients.

Although existing work extracted cancer information for numerous cancer types, we only created child frames for those types where any type-specific information was extracted. There were papers related to cancer types such as lung cancer [40][41], liver cancer [42] and ovarian cancer [43], but no frames were created for these types as they all extract the general cancer-related elements described in the 'CANCER DIAGNOSIS' frame.

Table 1 summarizes the frames created from the 79 selected papers, with corresponding per-element references. Supplementary Table 1 lists the per-frame references. The frame descriptions and the overall relationships between the frames are shown in Figure 2.

**Table 1: Frames (including frame elements) along with the associated articles**

| Frame | Frame Elements | References |
|---|---|---|
| **CANCER DIAGNOSIS** | NAME: cancer type | [44], [45], [42], [46], [47], [48], [49], [50], [51] |
| | ANATOMICAL SITE: the location description of the finding (including primary and metastatic sites) | [45], [52], [42], [53], [54], [55], [25], [27], [56], [57] |
| | HISTOLOGY: histological description (e.g. carcinoma) | [44], [52], [58], [55], [53], [54], [4], [27], [59], [43], [57] |
| | GRADE: appearance of the cancerous cells, can be frame with further information (GRADING VALUE) | [44], [52], [54], [4], [48], [27], [59], [60], [43], [61], [62] |
| | INVASION TYPE: the stage or level of invasion | [52] |
| | TUMOR BLOCK: tissue cores removed from regions of interest in paraffin-embedded tissues (e.g. 0.6 mm in diameter) | [52] |
| | TISSUE BANK: identifiers about location of tissue samples within an institution | [52] |
| | STATUS: whether confirmed, suspected and there is no evidence of finding (e.g. probable, definite, without) | [42], [57] |
| | RECURRENT STATUS: the value of recurrent status | [42], [63], [57] |
| | TEMPORAL INFORMATION: refers to information about time (e.g., year, month, and date, 2007-08-04) | [42], [57] |
| | SPECIMEN TYPE: the type of specimen involved in diagnosis | [53] |
| | LATERALITY: describes the side of a paired organ associated with origin of the primary cancer | [53], [54], [25], [48], [64], [27], [51] |
| | TUMOR SIZE: how large across the tumor is at its widest point (part of cancer staging) | [52], [53], [54], [25], [48], [65], [59], [60], [62], [57] |
| | TNM STAGE: cancer staging system, can be a separate frame with further information (TNM CLASSIFICATION) | [55], [53], [2], [3], [25], [66], [67], [60], [50], [40], [61] |
| | EXTENSION: direct extension of tumor | [53] |
| | UNCERTAINTY: used to differentiate clinical suspicions from conclusive findings (e.g., possible, likely) | [68], [57] |

| Frame | Frame Elements | References |
|---|---|---|
| | NEGATION: existence/negation of diagnosis (e.g., no, positive) | [68], [57] |
| | DISEASE: disease related concepts related to diagnosis (disease stage and severity) | [11], [25] |
| | DISEASE EXTENT: determine extent of disease (e.g., non-invasive, invasive, or metastatic) | [68], [69], [57] |
| | STAGE: the overall stage of cancer (e.g. Stage 0, Stage I) | [70], [52], [42], [54], [25], [69], [40], [43], [61], [57] |
| | EXISTENCE: existence description of the finding | [45] |
| | TEMPORAL MODIFIER: temporal modifiers of the finding | [45] |
| | ASSOCIATION: associations with other findings (may or may not be related to cancer) (e.g. causal, differential interpretation, and co-occurring) | [45] |
| **BREAST CANCER DIAGNOSIS** | STATUS OF CANCER TYPES: presence or absence of various types of breast cancer (e.g. ductal carcinoma in situ, invasive lobular carcinoma, atypia) | [47], [64], [49], [65], [24] |
| | RECEPTOR NAME: estrogen, progesterone, human epidermal growth factor 2 | [71], [72], [25] |
| | RECEPTOR STATUS: positive/negative | [72], [48], [66], [64], [69], [65], [60] |
| | RECEPTOR STATUS NEGATION: negation of status from surrounding text of receptor status mention | [72] |
| | RECEPTOR PERCENT: number of cells out of 100 that stained positive for a receptor | [48], [60] |
| | EXTRACAPSULAR AXILLARY NODAL EXTENSION STATUS: presence/absence of extracapsular extension in axillary lymph nodes | [64] |
| | ISOLATED CANCER CELLS IN LYMPH NODES STATUS: presence/absence of isolated cancer cells in sentinel lymph nodes | [64] |
| | MENOPAUSAL STATUS: status of menopause | [69] |

| Frame | Frame Elements | References |
|---|---|---|
| | SCARFF-BLOOM-RICHARDSON (SBR) STAGE -prognostic factor in breast cancer, associated with cell proliferation, also a consistent indicator of response to chemotherapy | [65] |
| | CONTRALATERAL EVENT: event of detecting a tumor in the opposite breast which was diagnosed more than 6 months following the detection of the first cancer | [51] |
| | MEDIASTINAL AND/OR STERNAL INVOLVEMENT: metastatic to sternum or mediastinum | [73] |
| **COLORECTAL CANCER DIAGNOSIS** | POLYPS TYPE: type of polyp present at the time of colonoscopy procedure (e.g., advanced conventional adenomas) | [59] |
| **BLADDER CANCER DIAGNOSIS** | INVASION STATUS: presence or absence of invasion | [44], [58], [4] |
| | DEPTH OF INVASION: measured from the basement membrane of epithelium from which the tumor is considered to arise, to the deepest point of invasion (e.g., superficial and muscle invasion) | [44], [58], [4], [50] |
| | MUSCULARIS PROPRIA: presence or absence of muscle in the specimen | [44], [58], [4] |
| | CARCINOMA IN SITU: statements regarding presence of carcinoma in situ, a group of abnormal cells | [58], [4], [67], [61] |
| **SKIN CANCER DIAGNOSIS** | CLARK LEVEL: described in separate frame (CLARK LEVEL) | [5] |
| | BRESLOW DEPTH: described in separate frame (BRESLOW DEPTH) | [5] |
| **PROSTATE CANCER DIAGNOSIS** | GLEASON SCORE: described in separate frame (GLEASON SCORE) | [5], [68], [3] |
| | PSA: prostate-specific antigen value | [3] |
| **COMORBIDITY DIAGNOSIS** | NAME: name of the comorbidity diagnosis (e.g. Liver cirrhosis) | [42] |
| | DIAGNOSIS STATUS: Whether confirmed, suspected and there is no evidence of finding (e.g. probable, definite, without) | [42] |

| Frame | Frame Elements | References |
|---|---|---|
| | CHILD-PUGH STAGING: e.g. Child-Pugh score: class A) | [42] |
| | TEMPORAL INFORMATION: refers to information about time | [42] |
| | REPORT TYPE: type of the report (e.g. Computed Tomography) | [42] |
| **FAMILY HISTORY** | FAMILY MEMBER: associated member | [74] |
| | DIAGNOSIS: associated cancer diagnosis | [74] |
| | RELATION: Relation between the family member and diagnosis | [74] |
| | NEGATION: negated mention of history | [74] |
| **KI-67 EXPRESSION** | KI-67 SCORE: a cancer antigen that is expressed during cell growth and division, but is absent in the cell resting phase | [6], [75], [48], [65], [62] |
| **GRADING VALUE** | SCALE: cancer is usually graded on a scale of 1-3 (lower number indicates cancer cells look more similar to normal cells) | [5], [52] |
| | TYPE: grading categories (e.g. Grade 1 – well differentiated) | [52], [54], [4], [43], [61] |
| **CLARK LEVEL** | VALUE: describes the level of anatomical invasion of the melanoma in the skin | [5] |
| **BRESLOW DEPTH** | VALUE: the distance between the upper layer of the epidermis and the deepest point of tumor penetration | [5] |
| **GLEASON SCORE** | SCORE: grading system used to determine the aggressiveness of prostate cancer | [5], [68], [76] |
| **ASSESSMENT FOR PROSTATE CANCER SURGERY** | GLEASON SCORE: described in separate frame (GLEASON SCORE) | [76] |
| **PRE-TREATMENT PROSTATE CANCER ASSESSMENT** | GLEASON SCORE: described in separate frame (GLEASON SCORE) | [77] |
| **ASSESSMENT FOR CANCER SURGERY** | TNM STAGE: cancer staging system, a separate frame with further information (TNM CLASSIFICATION) | [76] |

| Frame | Frame Elements | References |
|---|---|---|
| | MARGIN STATUS: status of surgical margin, described in separate frame (MARGIN) | [76] |
| **PRE-TREATMENT CANCER ASSESSMENT** | DOCUMENTATION: documentation within 6 months prior to initial treatment (e.g. treatment of prostate-specific antigen) | [77] |
| | PROCEDURE: performing diagnostic test, described in frame (CANCER PROCEDURE) | [77] |
| **MARGIN** | TYPE: type (e.g., anatomical, surgical) | [52] |
| | STATE/STATUS: if the inked margin of resection is found to contain tumor, it indicates that the surgeon inadvertently incised into the tumor, resulting in a positive surgical margin (e.g. positive surgical margin, negative surgical margin, not applicable or no explicit diagnosis provided) | [52], [76] |
| | DIMENSION: margin size | [52] |
| **ASSESSMENT FOR CANCER CARE** | SURGERY RELATED: described in separate frame (ASSESSMENT FOR CANCER SURGERY) | [76] |
| | PRE-TREATMENT RELATED: pretreatment process quality measures, described in frame (PRE-TREATMENT CANCER ASSESSMENT) | [77] |
| **PAIN IN PROSTATE CANCER PATIENTS** | MENTION: mention of pain related term | [78] |
| | PAIN SEVERITY: associated with a severity level from the four-tiered pain scale (no pain - category 0; some pain - category 1; controlled pain - category 2; severe pain - category 3) | [78] |
| | INTERNAL DATES: pain start and end dates | [78] |
| | LOCATION: body location of pain | [78] |
| | NEGATION: negated mentions of pain related terms | [78] |
| **CANCER PROCEDURE** | NAME: name of the procedure/test (e.g. chest x-ray) | [6], [79], [80], [75], [71], [81], [15], [1], [9], [82], [83], [60], [50], [43] |

| Frame | Frame Elements | References |
|---|---|---|
| | CODE TERMINOLOGY: clinical terminology used for the procedure mention | [52] |
| | CODE VALUE: value of the terminology code | [52], [1] |
| | INSTITUTION: institution where the procedure was performed | [52] |
| | NEGATION: whether the mention is negated | [52], [77] |
| | MENTION: words related to any procedure term (e.g. flex sig, guaiac card) | [75], [52], [15], [84], [77], [43] |
| | MARGIN: usually the rim of normal tissue taken removed during or after procedure (surgical margin) | [52] |
| | ANATOMICAL SITE: part of body procedure targets (e.g., breast) | [79], [52] |
| | TEMPORAL INFORMATION: time and date descriptors (e.g., "colonoscopy in 2005", "flexible sigmoidoscopy 5 years ago), date of completion | [52], [15], [1], [10] |
| | STATUS: procedure or treatment status (e.g., refused, declined, scheduled, planned, completed, reported vs not reported) | [15], [42], [1], [82] |
| | MODIFIER: negation and other modifiers that change the status of procedure (e.g., "no", "never") | [15] |
| **TUMOR DESCRIPTION** | ANATOMICAL SITE: anatomic locations (e.g., "segment 5" or "left lobe") with attributes (Liver, Non Liver), target location (e.g., liver and segment #7) as well as non-target location (e.g., breast) | [13], [8], [52], [85], [42], [25], [9], [23], [26], [41] |
| | LATERALITY: side of a paired organ associated with origin of the primary tumor | [25] |
| | TYPE: primary/metastatic | [25] |
| | STATUS: benign or malignancy status along with diagnostic information such as 'suggestive of cyst' | [85], [86], [41], [57] |
| | HISTOLOGY: morphologic type of the tumor | [52], [9] |
| | STAGE: tumor stage | [8], [68] |

| Frame | Frame Elements | References |
|---|---|---|
| | GRADE: appearance of the cancerous cells | [8], [52], [9], [48], [87], [26] |
| | INVASION: whether or not more than 50% of an organ is invaded | [52], [88] |
| | SIZE: quantitative size of tumor (e.g., 2.2 x 2.0 cm), diameter/volume of the tumor, including unit (e.g., 1 cm, 0.3 x 0.5 x 0.7 cm) | [13], [52], [88], [85], [42], [25], [48], [41] |
| | SIZE TYPE: radiological/pathological | [25] |
| | NEGATION: indicator to some negation of a tumor reference (e.g., no) | [85], [86], [41] |
| | COUNT: number of tumor/nodule references (e.g., two or multiple) | [13], [88], [85] |
| | TUMOR REFERENCE: a radiologic artifact that may reference a tumor (e.g., lesion or focal density) | [88], [85] |
| | MENTION: tumor major object (e.g., tumor, lesion, mass, and nodule) | [13], [42], [47], [49], [41] |
| | QUANTIFIER: one, two, three, several | [42] |
| | TEMPORAL INFORMATION: refers to information about time (e.g., year, month, and date, 2007-08-04) | [52], [42], [86] |
| | NON-TUMOR SIZE ITEMS: LeVeen needle, which is used in RFA treatment | [42] |
| | STATUS: this indicates the final overall tumor status (e.g., regression, stable, progression, irrelevant) | [89], [41] |
| | METASTATIC STATUS INDICATORS: phrases denoting a metastatic tumor | [9], [90] |
| | MAGNITUDE: indicates the qualitative extent of change, if any (e.g., mild, moderate, marked) | [89] |
| | SIGNIFICANCE: indicates the subjective clinical significance of change, if any (e.g., uncertain, possible, probable) | [89] |
| | TRACK ASSIGNMENT: Fleischner Society based surveillance track assignment for patients who received LCS LDCT (guidelines about follow-up procedure depending on the nodule size) | [13] |

| Frame | Frame Elements | References |
|---|---|---|
| | PROCEDURE: method for obtaining the tumor, can be a separate frame (CANCER PROCEDURE) | [13], [9], [50] |
| | TNM STAGE: cancer staging system, a separate frame with further information (TNM CLASSIFICATION) | [52], [87], [50] |
| | CLOCK-FACE: clock-face location of the tumor | [25] |
| **BLADDER TUMOR DESCRIPTION** | MUSCLE INFORMATION: presence or absence of muscle, for those that mentioned muscle, rates of muscle presence in the surgical specimen | [8], [87] |
| | CARCINOMA IN SITU: statements regarding presence of carcinoma in situ | [87] |
| **CANCER TREATMENT** | TYPE: drug, procedure, radiation, etc. | [25], [10], [43] |
| | PATIENT CENTERED OUTCOMES: outcomes as interpreted and documented by clinicians following a patient's cancer treatment along with their semantic context such as negations | [91] |
| | PATIENT REPORTED OUTCOMES: symptoms experienced by patients during cancer treatment along with negations | [92] |
| | SITE: metastatic site of cancer where the treatment was targeted | [10] |
| **TNM CLASSIFICATION** | TUMOR SIZE: diameter/volume of the tumor, including unit (e.g., 3-4 mm) | [6], [67], [60] |
| | REGIONAL LYMPH NODES INVOLVED: regional lymph nodes information used to detect staging | [6], [93] |
| | METASTASIS: whether the tumor has invaded the nearby tissues | [6], [50] |
| | DISTANT METASTASIS: whether tumor has spread from the original (primary) tumor to distant organs or distant lymph nodes | [6], [93] |
| | GRADE: appearance of the cancerous cells | [6], [67], [60] |
| | STAGING FACTORS: factors relevant to staging and used to calculate the TNM stage | [93] |

| Frame | Frame Elements | References |
|---|---|---|
| | STAGE: AJCC stage designation (e.g., T1, N1b) | [2], [3], [25], [66], [60], [50], [40], [61] |
| | TIMING: used to indicate if the staging is clinical or pathological as per the rules of the AJCC manual | [2], [3], [25], [66], [50], [40] |
| | NEGATION: any negated mention | [2] |
| | TEMPORALITY: capture historical or future mentions that do not necessarily represent current mentions valid at the point in time when the mention was stated at the patient record | [2] |
| | SUBJECT INVOLVED: capture TNM mentions related to family relatives or others | [2] |
| **PERFORMANCE STATUS** | NAME: name of the performance status scale, used as a prognostic tool (e.g. ECOG, Karnofsky) | [55], [40] |
| | SCORE: measure of functional status (ECOG score ranges from 0-5, whereas Karnofsky Scale' ranges from 0-100) | [40] |
| **DIAGNOSTIC PROCEDURE** | PROCEDURE RESULT: result of a procedure mentioned in the report (e.g., Calcifications, hyperplasia etc.) | [79] |
| | TEST RESULT: result of the test (e.g. positive vs negative) | [14], [82], [94], [83], [77] |
| **THERAPEUTIC PROCEDURE** | TYPE: treatment type (e.g., RFA and TACE) | [42], [60], [50] |
| | LINE OF THERAPY: initial treatment is referred to as first-line treatment or first-line therapy, however, a second-line treatment may be suggested later | [50] |
| | THERAPY DOSE: the total about of treatment (e.g., radiation) the patient is exposed to (e.g. radiation therapy dose) | [10] |
| | TOXICITIES: various toxicities related to cancer treatment therapy along with negation and certainty | [10] |
| **CANCER FINDING** | PROCEDURE: name of the associated cancer procedure (e.g., Breast Core biopsy), can be a separate frame (CANCER PROCEDURE) | [7], [79], [71], [81], [11] |
| | FINDING TYPE: type of the finding (e.g., negative, normal, positive, possible, | [80] |

| Frame | Frame Elements | References |
|---|---|---|
| | probably, history, mild, stable, improved, or recommendation) | |
| | FINDING MODIFIER: modifying words from within the report (e.g., type: mild, modifiers: tortuous) | [80] |
| | LATERALITY: location or sidedness of the finding (e.g., "left", "right," "both", or "bilateral") | [79] |
| | BODY PARTS: body organs on which the finding is reported | [11] |
| **PATHOLOGY FINDING** | POSITIVE LYMPH NODES NUMBER: number of positive lymph nodes mentioned in the finding | [71], [48], [65] |
| | POSITIVE LYMPH NODES STATUS: presence or absence of positive lymph nodes | [64] |
| | LYMPH NODES REMOVED: number of lymph nodes removed | [48] |
| | NUCLEAR GRADE: describes how closely the nuclei of cancer cells look like the nuclei of normal cells | [71] |
| | PLOIDY: refers to amount of DNA the cancer cells contain | [71] |
| | QUALITATIVE S-PHASE: indicator of tumor growth rate | [71] |
| | BIOMARKER: name of the biomarker | [12], [40] |
| | BIOMARKER TEST RESULTS / MUTATION STATUS: positive or negative for biomarkers such as ALK and EGFR | [12], [40], [55], [65] |
| | SPECIMEN: specimens on which findings are described (e.g. core biopsies or organ portions) | [66] |
| | STATUS OF INVASION: presence or absence of invasion such as blood vessel invasion) | [64] |
| **IMAGING FINDING** | CALCIFICATION: presence or absence of calcification | [7], [95], [96], [24] |
| | MASS: presence or absence of mass | [7], [96], [24] |
| | ARCHITECTURAL DISTORTION: whether architectural distortion is present or absent | [7], [96], [24] |
| | CYSTS: whether cyst is present or absent | [7], [95] |

| Frame | Frame Elements | References |
|---|---|---|
| | NME: whether non-mass enhancement is present or absent | [7] |
| | FOCUS: whether focus is present or not in imaging such as breast magnetic resonance imaging | [7] |
| | RECOMMENDATION: follow-up recommendation | [81] |
| | CALCIFICATION CHARACTERISTICS: characteristics of calcification | [96] |
| | MUTATION CHARACTERISTICS: imaging descriptors/features to distinguish wild-type vs mutated patients (e.g. distinguish wild-type and KRAS mutations) | [97] |
| **BREAST IMAGING FINDING** | IMPLANTS: whether breast implants are present or absent | [7] |
| | ASYMMETRY: whether asymmetry is present or absent | [7], [96], [24] |
| | BIRADS CATEGORY: refers to the number stated for each breast and laterality (e.g. 1 for negative category) | [7], [81] |
| | DENSITY: breast density or breast composition | [96], [24] |

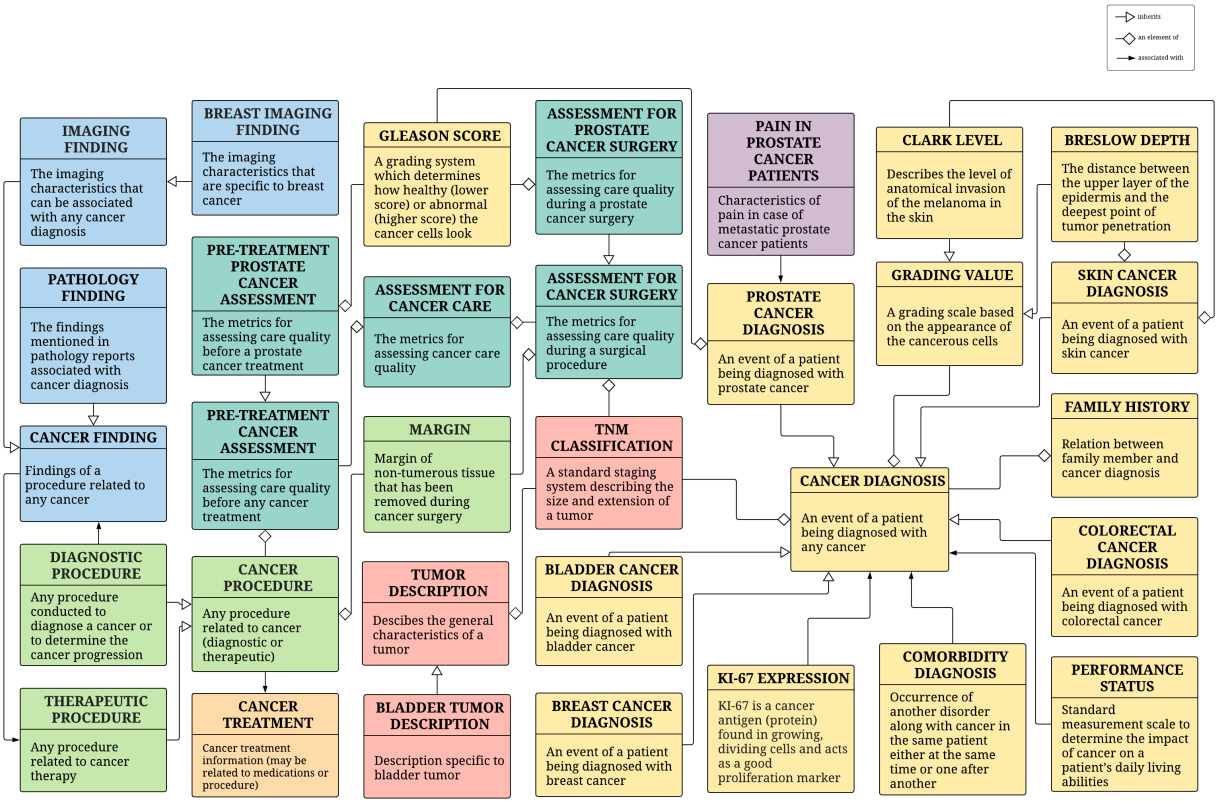

**Figure 2: Frames and their relations as expressed in literature**

**Most Extracted Frames:**

As demonstrated in Table 1, 'CANCER DIAGNOSIS' is the most referenced frame overall, i.e. it has the highest number of associated articles (*n*=36). This was expected as most efforts focused on extracting general information related to diagnosis of cancer. However, this does not necessarily mean that a paper included in 'CANCER DIAGNOSIS' frame extracted all the elements present in that frame. We observe that a subset of papers extracting some specific elements of this frame also extracted elements present in one of its child frames. For instance, a breast cancer related article by Breischneider et al. [60] extracted general cancer data elements such as TNM CLASSIFICATION and GRADE present in 'CANCER DIAGNOSIS' frame whereas it also extracted RECEPTOR STATUS and RECEPTOR PERCENT elements described in the 'BREAST CANCER DIAGNOSIS' frame. Thus, some papers extract only a subset of elements present in one or more frames and not all of the elements related to a single frame. In terms of the total number of articles associated with a frame, the 'CANCER DIAGNOSIS' frame is followed by 'TUMOR DESCRIPTION' (*n*=21) and CANCER PROCEDURE' (*n*=19) frames.

**Most Extracted Frame Elements**

We also note the distribution of papers at the frame element level. This provides an idea about the important clinical attributes found in EHR text related to cancer that interest researchers. Table 2 highlights the 10 most extracted frame elements along with the number of referenced articles.

**Table 2: Most extracted frame elements along with the number of referenced articles**

| Frame Element | Number of articles extracting this element |
|---|---|
| GRADE | 19 |
| ANATOMICAL SITE | 18 |
| TUMOR SIZE | 17 |
| TNM CLASSIFICATION | 16 |
| CANCER PROCEDURE NAME | 16 |
| STAGE RELATED (Overall + TNM) | 15 |
| RECEPTOR/BIOMARKER RELATED | 13 |
| HISTOLOGY | 13 |
| STATUS (tumor and cancer related) | 9 |
| LATERALITY | 8 |

**Papers Extracting Multiple Frames**

Many eligible papers extracted varied types of cancer information corresponding to different frames. The NLP system developed by Savova et al. [25] extracted various cancer phenotypes from EHRs, which were then used to generate summaries containing both cancer and tumor characteristics. A similar trend was observed for other papers including [42] and [52], where information related to multiple frames were extracted. Notably, some frames (particularly the ones which are not directly associated) contain a few common elements. For example, TUMOR SIZE element is present in both 'CANCER DIAGNOSIS' and 'TUMOR DESCRIPTION' frames as size is one of the characteristics that can be associated with both tumor and cancer diagnosis (cancer being linked to tumor and tumor characteristics). Thus, assigning any paper to such a common frame element (e.g., TUMOR SIZE) was primarily based upon whether the paper dealt specifically with tumor description or cancer diagnosis. Table 3 shows a detailed overview of some of the noteworthy papers extracting information that could be represented using at least four different cancer frames.

**Table 3: Papers extracting varied frames**

| Paper | Number of frames extracted | Frames extracted | Number of frame elements extracted |
|---|---|---|---|
| Cary et al. [50] | 6 | CANCER DIAGNOSIS, BLADDER CANCER DIAGNOSIS, TNM CLASSIFICATION, TUMOR DESCRIPTION, CANCER PROCEDURE, THERAPEUTIC PROCEDURE | 11 |
| Napolitano et al. [5] | 6 | PROSTATE CANCER DIAGNOSIS, SKIN CANCER DIAGNOSIS, GLEASON SCORE, GRADING VALUE, CLARK LEVEL, BRESLOW DEPTH | 7 |
| Coden et al. [52] | 5 | CANCER PROCEDURE, CANCER DIAGNOSIS, TUMOR DESCRIPTION, GRADING VALUE, MARGIN | 28 |
| Ping et al. [42] | 5 | CANCER PROCEDURE, THERAPEUTIC PROCEDURE, TUMOR DESCRIPTION, CANCER DIAGNOSIS, COMORBIDITY DIAGNOSIS | 19 |
| Savova et al. [25] | 5 | TUMOR DESCRIPTION, CANCER TREATMENT, CANCER DIAGNOSIS, BREAST CANCER | 16 |

| Paper | Number of frames extracted | Frames extracted | Number of frame elements extracted |
|---|---|---|---|
| | | DIAGNOSIS, TNM CLASSIFICATION | |
| Tang et al. [48] | 5 | CANCER DIAGNOSIS, BREAST CANCER DIAGNOSIS, PATHOLOGY FINDING, KI-67 EXPRESSION, TUMOR DESCRIPTION | 11 |
| Breischneider et al. [60] | 5 | CANCER DIAGNOSIS, BREAST CANCER DIAGNOSIS, TNM CLASSIFICATION, CANCER PROCEDURE, THERAPEUTIC PROCEDURE | 10 |
| Xu et al. [71] | 4 | CANCER FINDING, PATHOLOGY FINDING, CANCER PROCEDURE, BREAST CANCER DIAGNOSIS | 7 |
| Thiebaut et al. [65] | 4 | CANCER DIAGNOSIS, BREAST CANCER DIAGNOSIS, KI-67 EXPRESSION, PATHOLOGY FINDING | 7 |
| Schroeck et al. [61] | 4 | CANCER DIAGNOSIS, BLADDER CANCER DIAGNOSIS, TNM CLASSIFICATION, GRADING VALUE | 6 |

**Significant Cancer Types Based on Elements Extracted**

Although a large proportion of the papers reviewed extracted general cancer information from EHRs using NLP, there are some papers which focused on extracting cancer-type specific information. The most notable cancer type was breast cancer, with the highest number of associated articles (*n=14*), followed by bladder cancer (*n=6*) and prostate cancer (*n=4*).

The frame elements for the three most commonly extracted cancer types based on the number of articles are summarized in Table 4. There may be more articles related to each of these cancer types, but the number here includes those that extract only type specific information (e.g., information specific to breast cancer instead of general cancer information extracted only from the notes of breast cancer patients).

**Table 4: Most extracted cancer types**

| Cancer type | Important frame elements specific to this cancer | Number of associated papers |
|---|---|---|
| Breast Cancer | STATUS OF CANCER TYPES, RECEPTOR NAME (mainly Estrogen, progesterone, human epidermal growth factor 2), RECEPTOR STATUS, RECEPTOR PERCENT, RECEPTOR STATUS NEGATION, SCARFF-BLOOM-RICHARDSON STAGE, CONTRALATERAL EVENT | 14 |
| Bladder Cancer | INVASION STATUS, DEPTH OF INVASION, MUSCULARIS PROPRIA, CARCINOMA IN SITU | 6 |
| Prostate Cancer | GLEASON SCORE, PSA | 4 |

We also note in Table 1 that the 'PROSTATE CANCER DIAGNOSIS' frame has been associated with 'PAIN IN PROSTATE CANCER PATIENTS' frame, all of whose elements were extracted from clinical text in a case of metastatic prostate cancer patients for identifying unknown pain phenotypes [78]. Napolitano et al. [5] extracted 'BRESLOW DEPTH' and 'CLARK LEVEL', both of which are grading values for skin cancer, as well as 'GLEASON SCORE' from pathology reports.

**Frames Related to Cancer Quality Measures**

A few papers focused on automatically identifying quality measures information for assessing the handling of cancer patients such as grouping patients for clinical trials and making decisions about their treatment plans. The frames created for these quality measures, however, are general-purpose (largely different types of patient assessments) and could be used for other applications as well. D'avolio et al. [76] specifically worked on extracting such measures like MARGIN STATUS for patients who underwent prostatectomies. Another paper by Hernandez-Boussard et al. [77] extracted information about pre-treatment quality metrics. In context to assessing quality for cancer care, the frame 'ASSESSMENT FOR CANCER CARE' contains two elements, ASSESSMENT FOR SURGERY and PRE-TREATMENT CANCER ASSESSMENT, each of which further describe their individual elements (shown in Table 1). We observe in the table that besides GLEASON SCORE and surgical MARGIN STATUS which is already stated above, 'TNM CLASSIFICATION' is also identified by the College of American Pathologists (CAP) as one of the three Category I measures [76] for assessing quality.

**Frames Related to Cancer Screening**

Some articles applied NLP techniques to extract colonoscopy testing or colorectal cancer (CRC) screening related information such as test MENTION, TEMPORAL INFORMATION (e.g. 'colonoscopy in 2005'), STATUS (e.g. 'refused', 'scheduled') and NEGATION (e.g., negation modifiers such as 'no' and 'never') of the tests [14][15][1]. Since the screening tests facilitate the cancer diagnosis process, we have represented the information using the 'DIAGNOSTIC PROCEDURE' frame (Table 1), which inherits from the 'CANCER PROCEDURE' frame (Figure 2).

Another study by Ritzwoller et al. [13] extracted general characteristics of lung nodules (e.g. nodule size) from radiology reports that the providers submitted to a Centers for Medicare and Medicaid Services (CMS)-approved registry following any Low-Dose CT Lung cancer screening procedure. As represented in Table 1, all nodule related information are captured in the 'TUMOR DESCRIPTION' frame.

**Frames Related to Image Findings**

While a majority of papers identified information from pathology findings (which usually contain detailed tumor-related information), a few studies attempted to automatically extract data elements from imaging findings (e.g. mammography reports) [7][81][96][24]. Interestingly, these findings were mostly related to breast cancer screening. For example, Lacson et al. [7] extracted elements such as CALCIFICATION, MASS, IMPLANTS, BIRADS CATEGORY, CYSTS, etc. with the aim to use these extracted elements for populating a breast cancer screening registry. He et al. [24] employed deep learning and NLP methods to extract similar information such as BREAST DENSITY, MASS, ARCHITECTURAL DISTORTION etc. from mammographic findings for identifying high risk patients and patients for whom biopsy is recommended. We have represented all the imaging related data elements in the 'IMAGING FINDING' frame and its subframe, 'BREAST IMAGE FINDING'.

**Other Important Frame Elements**

Apart from the above mentioned frames and their respective elements, some of the other important data elements that multiple researchers (at least 3) were concerned about were: TEST RESULT in the 'DIAGNOSTIC PROCEDURE' frame [14][82][94][77]; TEMPORAL INFORMATION [15][42][1][10] and STATUS [15][42][1][82] in the 'CANCER PROCEDURE' frame; CALCIFICATION, MASS, ASYMMETRY, ARCHITECTURAL DISTORTION in the 'IMAGING FINDING' frame [7][96][24]; TYPE in the 'THERAPEUTIC PROCEDURE' frame [42][60][50]; and KI-67 SCORE in the 'KI-67 EXPRESSION' frame [6][75][48][65].

Two of the studies we reviewed extracted a patient's performance status information. One such performance measure is the Eastern Cooperative Oncology Group (ECOG) Scale of Performance Status, extracted by Herath et al. [55] and Najafabadipour et al. [40], whereas Karnofsky measure was extracted additionally by Najafabadipour et al. [40] from clinical narratives.

**DISCUSSION**

Our scoping review provides a detailed review of current research in the cancer information extraction domain from unstructured EHR sources using NLP. 173 papers were identified as relevant to clinical NLP for cancer. Of these, 79 were included in the frame structure (Table 1 and Figure 2). Appropriate interpretation of cancer information extracted as frames requires context, specifically tying the frames to the data source (EHR free text notes, in our case). We found that many papers use non-EHR sources for extracting cancer information (e.g., from biomedical literature), and these likely would have had entirely different frame structures (e.g., less patient-focused, more general knowledge-focused, more genomic information).

Our review demonstrates evidence of the growing interest among NLP researchers in automatically extracting cancer related information from clinical text. The following points synthesize our key findings:

The first key finding is the redundancy of certain information types (e.g., 19 papers extracted cancer grade information). This speaks to a tremendous replication of effort in these areas (Table 2), as state-of-the-art NLP systems require medium to large manually-annotated training sets to build high-performance machine learning models. Yet, as far as we can tell, the papers describe systems developed exclusively on local data and do not involve the sharing of data or models, forcing future researchers and clinicians in need of these cancer information types to "reinvent the wheel", replicating a time-consuming process requiring substantial NLP expertise. Pilot projects are underway to meet some of the needs of cancer researchers requiring such systems [25] [57], but the breadth of overlap (Table 1) suggests that without a more widely-available open resource, continued effort duplication will continue. Even more daunting are the cancer information needs that go unfulfilled because the costs of NLP projects are too high. This review demonstrates that there is a tremendous need for a more general-purpose cancer NLP resource.

The second key finding is that, given the fact that so few papers focus on more than a few information types, the decision on which cancer types to extract for a project is clearly an ad hoc process. The genesis of most of these papers was the need to support a given research project or clinical use case. This leads, for example, to cancer-specific diagnostic information being performed only for bladder, breast, colorectal, prostate, and skin cancer. Missing are common cancers such as lung cancer, lymphoma, kidney cancer, and leukemia. Each of these have information types specific to that form of cancer (e.g., certain blood tests for leukemia, or a blood marrow biopsy result). Given, again, the high effort and cost that can be associated with an NLP project, this collection of ad hoc projects—with both sizable overlap and yet key gaps—would be better served by a more general-purpose effort. This effort would identify widely-needed cancer information types and create an NLP system that supports a broad range of use cases. It would be impossible to cover all the use cases required by the 79 papers within this study, there are few specific, recurring information needs (Table 1).

**Impact of Frame Semantics**

This review organized cancer information according to the theory of frame semantics. As one can see from Figure 2, frames are an intuitive means of organizing information, not unlike entity-relationship and class diagrams from computer science. Frame semantics, however, is a fundamentally linguistic theory, as already described. This largely helps to minimize discrepancies between possible frame representations, though we make no claim the one presented here is the only or 'best' method for representing cancer information. There are certainly no shortage of ontologies that already exist to describe cancer information. But given the prominence of frames in general NLP methods, it is nonetheless beneficial to consider what a set of frames for cancer would look like when targeted toward NLP for EHR notes. Further, we did not find evidence, during the course of this review, of any emerging standard for organizing information extracted by NLP for cancer.

We consider this work, then, as the first step in an iterative process to derive an authoritative set of frames for cancer NLP targeted toward EHR notes. More investigation is certainly needed. Notably, frame structures need to stand up to the test of EHR text itself. Frame semantics asserts that the information (the elements) within a frame are often found together within some reasonable context (e.g., a cancer's GRADE and ANATOMICAL SITE can be found within the same sentence, if not the same clause). So for the proposed frames, this needs to be tested for frames to be used as a definitive NLP target representation.

Regardless of the utility in the proposed frames for actual NLP development, the organization of this information into frame structures is nonetheless useful. Table 1 contains over 160 elements, and any attempt to describe all of these information types without some intermediate structure (a frame that contains some set of elements) would be difficult to interpret. For instance, if a future researcher were interested in several specific information types related to cancer therapy procedures, finding the 'THERAPEUTIC PROCEDURE' frame will allow access to papers related to the information types organized under this frame (e.g., THERAPY DOSE and TOXICITIES information).

**LIMITATIONS**

This work has several limitations. The first limitation of this review is that, given the rapid pace of NLP development and publication, it is likely that papers meeting inclusion criteria were omitted. Specifically, papers published starting in late 2018 were likely missed. Second, although we have tried to accurately interpret and represent all the possible information types extracted in the literature as frames, there might be a few which are not captured or mis-represented in our final frame list. Finally, as already discussed, there may be inconsistencies in assigning the frames and associated elements across all the papers.

**CONCLUSIONS**

Our scoping review provides a detailed overview of the current research in the cancer information extraction domain from unstructured EHR sources using NLP. We conducted the review from a frame semantic perspective, described various frames along with their elements as well as examined the relations between frames. Since many researchers are trying to extract similar frames or frame elements (though not always using the language of frame semantics), this review can help developers of a general-purpose cancer frame resource and NLP system that would extract a broad range of important cancer information types.